\documentclass[12pt]{article}
\usepackage{amsfonts,amssymb,amsmath,amscd,theorem,oldgerm,epsfig,units}
\usepackage[mathscr]{eucal}

\newtheorem{lemma}{Lemma}[section]
\newtheorem{proposition}[lemma]{Proposition}
\newtheorem{remark}[lemma]{Remark}

\newtheorem{theorem}[lemma]{Theorem}

\newtheorem{definition}[lemma]{Definition}
\newtheorem{corollary}[lemma]{Corollary}
\newtheorem{claim}[lemma]{Claim}

\newcommand{\BProof}{{\it Proof.} \r}
\newcommand{\EProof}{\hfill \blacksquare}

\def\ome {\omega}

\def\XI{\xi}

\def\lam{\lambda}
\def\lto{\longrightarrow}

\def\iso{\backsimeq}

\def\R{{\mathbb R}}
\def\Q{{\mathbb Q}}
\def\N{{\mathbb N}}
\def\Z{{\mathbb Z}}
\def\C{{\mathbb C}}
\def\Lam{\Lambda}
\def\Lm{\Lambda^*}

\def\GL{\mathrm{GL}}

\def\GZ{\mathrm{SL}_2(\Z)}
\def\G{\mathrm{\Gamma}}

\def\AT{\mathbb{T}}      % algebraic standard torus

\def\S0{S}

\def\F{C^{\infty}({\bf T})}
\def\T{{\bf T}}
\def\Td{\T^\vee} % the lattice of characters on T

\def\Irr{\mathrm{Irr}({\cal A} _\hbar)}
\def\A{\cal A}
\def\Ad{ {\cal A} _\hbar}

\def\h{ \hbar }
\def\H{{\mathscr H}}
\def\Hh{{\H}_\h}
\def\W{{\mathrm W}}

\def\rhoh{\rho _{_\h}}

\def\Pih{\pi_{_\hbar}}
\def\Pi{\pi}
\def\rhop {\rho_{_{\mathrm p}}}

\def\P1{\mathbb{P}^1}

\def\i_XI{i_{_{\XI}}}

\def\p_XI{p_{_{\XI}}}

\def\sub{\subset}

\def\Hom{\mathrm{Hom}}

\def\cent{Z}
\def\rev{\quad}

\def\r{\;}

\def\q{{\bf{q}}} % the Gama equiariant map between Irr and Q_h
\def\f{\mathrm{f}}
\def\2n{\mathrm{2n}}
\def\n{{\mathrm n}}
\def\m{{\mathrm m}}
\def\y{{\mathrm y}}
\def\x{{\mathrm x}}

\begin{document}

\title{\bf \small THE TWO DIMENSIONAL HANNAY-BERRY MODEL}
\author{\small\it SHAMGAR GUREVICH AND RONNY HADANI}

\date{}

\maketitle

\bigskip

\begin{abstract}
The main goal of this paper is to construct the Hannay-Berry model
of quantum mechanics, on a two dimensional symplectic torus. We
construct a simultaneous quantization of the algebra of functions
and the linear symplectic group $\G =$ SL$_2 (\Z)$. We obtain the
quantization via an action of $\G$ on the set of equivalence
classes of irreducible representations of Rieffel`s quantum torus
$\Ad$. For $\h \in \Q$ this action has a unique fixed point. This
gives a canonical projective equivariant quantization. There
exists a Hilbert space on which both $\G$ and $\Ad$ act
equivariantly. Combined with the fact that every projective
representation of $\G$  can be lifted to a linear representation,
we also obtain linear equivariant quantization.
\end{abstract}

\bigskip

\numberwithin{equation}{subsection}

\setcounter{section}{-1}

\section{Introduction}
\subsection{Motivation}
In the paper ``{\it Quantization of linear maps on the torus -
Fresnel diffraction by a periodic grating}'', published in 1980
(cf. \cite{HB}), the physicists  J. Hannay and M.V. Berry explore
a model for quantum mechanics on the 2-dimensional torus. Hannay
and Berry suggested to quantize simultaneously the functions on
the torus and the linear symplectic group $\G = $ SL$_2(\Z)$. They
found (cf. \cite{HB},\cite{Me}) that the theta subgroup $\G
_\Theta \sub \G$ is the largest that one can  quantize and asked
(cf. \cite{HB},\cite{Me}) whether the quantization of $\G$ satisfy
a multiplicativity property (i.e., is a linear representation of
the group). In this paper we want to \textit{construct} the
Hannay-Berry's model for the bigger group of symmetries, i.e., the
whole symplectic group $\G$. The central \textit{question} is
whether there exists a Hilbert space on which a deformation of the
algebra of functions and the linear symplectic group $\G$ both act
in a compatible way.
\subsection{Results}
In this paper we give an \textit{affirmative} answer to the
existence of the quantization procedure. We show a construction
(Theorem \ref{Cer}, Corollary \ref{prq} and Theorem \ref{lin}) of
the canonical equivariant quantization procedure for rational
Planck constants. It is \textit{unique} as a projective
quantization (see definitions below).  We show that the projective
representation of $\G$ can be lifted in exactly 12 different ways
to a linear representation (to obey the multiplicativity
property). These are the first examples of such equivariant
quantization for the whole symplectic group $\G$. Our construction
slightly \textit{improves} the known constructions \cite{HB, Me,
KR1} for which the group of quantizable elements is $\G_\Theta
\subset \G$ and gives a \textit{positive} answer to the
Hannay-Berry question on the linearization of the projective
representation of the group of quantizable elements. (cf.
\cite{HB}, \cite{Me}). Previously it was shown by Mezzadri and
Kurlberg-Rudnick (cf. \cite{Me}, \cite{KR1}) that one can
construct an equivariant quantization for the theta subgroup, in
case when the Planck constant is of the form $\h = \frac{1}{N},\;
N\in \N $.
\subsubsection{Classical torus}
Let $(\T,\ome)$ be the two dimensional symplectic torus. Together
with its linear symplectomorphisms $\G \iso \GZ$ it serves as a
simple model of classical mechanics (a compact version of the
phase space of the harmonic oscillator). More precisely, let $\T=
\W/ \Lambda$ where $\W$ is a two dimensional real vector space,
i.e.,  $\W \simeq \R^2$ and $\Lambda$ is a rank two lattice in
$\W$, i.e., $\Lambda \simeq \Z^2$. We obtain the symplectic form
on $\T$ by taking a non-degenerate symplectic form on $\W$:
$$\ome: \W \times \W \lto \R.$$
We require $\ome$ to be integral, namely $\ome : \Lambda \times
\Lambda \lto
\Z$ and normalized, i.e., Vol$(\T) =1$.\\\\
Let ${\mathrm{Sp}}(\W, \ome)$ be the group of linear
symplectomorphisms, i.e., $\mathrm{Sp}(\W,\ome) \simeq
\mathrm{SL}_2(\R)$. Consider the subgroup $\G \subset
\mathrm{Sp}(\W,\ome)$ of elements that preserve the lattice
$\Lambda$, i.e., $\G (\Lambda) \subseteq \Lambda $. Then $\G
\simeq \GZ$. The
subgroup $\G$ is the group of linear symplectomorphisms of $\T$.\\
We denote by $\Lm \subseteq \W^*$ the dual lattice:
$$
\Lm := \{ \xi \in \W^* | \r\r \xi (\Lambda) \subset \Z \}.
$$
The lattice $\Lm$ is identified with the lattice $\Td :=
\Hom(\T,\C^*)$ of characters of $\T$ by the following map:
\begin{equation*}
  \xi \in \Lm \longmapsto e^{2 \pi i <\xi, \cdot>} \in \r \Td.
\end{equation*}
The form $\ome$ allows us to identify the vector spaces $\W$ and
$\W^*$. For simplicity we will denote the induced form on $\W^*$
also by $\ome$.
\subsubsection{Equivariant quantization of the torus}
We will construct a particular type of quantization procedure for
the functions. Moreover this quantization will be equivariant with
respect to the action of the ``classical symmetries'' $\G$:
\begin{definition} By \textit{Weyl quantization} of ${\cal A}$ we mean a family of
$\C$-linear, $*-$ morphisms $\Pih:{\cal A} \lto $ End$(\H _\h), \r
\h \in \R$, where $\H _\h$ is a Hilbert space,  s.t. the following
property holds:
\begin{equation*}
    \pi _{_\h} (\xi+\eta) = e^{\pi i \h w(\xi,\eta)} \pi _{_\h}(\xi) \pi
_{_\h}(\eta)
\end{equation*}
for all $\xi,\eta \in \Lam ^*$ and $\h \in \R$.
\end{definition}
This type of quantization procedure  will obey the ``usual''
properties (cf. \cite{D4}):
\begin{eqnarray*}
|| \pi _{_\h}(fg) - \pi _{_\h}(f) \pi _{_\h}(g)||_{_{\H _\h}} &
\lto &  0, \rev   as \r \h  \to  0,
\\
||\frac{i}{\h}[\pi _{_\h}(f),\pi _{_\h}(g)]- \pi
_{_\h}(\{f,g\})||_{_{\H _\h}} & \lto &  0, \rev   as \r \h \to 0.
\end{eqnarray*}
where $\{,\}$ is the Poisson brackets on functions.
\begin{definition} By \textit{equivariant quantization} of $\T$ we mean a quantization of $\cal A$ with additional maps
$\rho _{_\h} : \G \lto  \mathrm{U} (\H _\h)$ s.t. the following
equivariant property $($called {\it Egorov's identity}$)$ holds:
\begin{equation}\label{eqp}
{\rho _{_\h}  (B)}^{-1} \pi _{_\h}(f) \rho _{_\h} (B) = \pi
_{_\h}(f\circ B)
\end{equation}
for all $ \h \in \R , \r f \in {\cal A}$ and $B \in \G$. Here
$\mathrm{U} (\H _\h)$ is the group of unitary operators on $\H
_\h$. If  $(\rho _{_\h} , \H _\h) $ is a projective
$($respectively linear$)$ representation of the group $\G$ then we
call the quantization \textit{projective} $($respectively
\textit{linear}$)$.
\end{definition}
The idea of the construction is as follows: We use a "deformation"
of the algebra $\cal A$ of functions on $\T$. We define an algebra
$\Ad$, usually called the two dimensional non-commutative torus
(cf. \cite{Ri}). If $\h = \frac{M}{N} \in \Q$, then we will see
that all irreducible representations of $\Ad$ have dimension $N$.
We denote by $\Irr$ the set of equivalence classes of irreducible
algebraic representations of the quantized algebra. We will see
that $\Irr$ is a set "equivalent" to a torus.
\\
\\
The group $\G$ naturally acts on a quantized algebra $\Ad$ and
hence on the set $\Irr$. Let $\h = \frac{M}{N}$ with
$\mathrm{gcd}(M,N) =1$. The following holds:
\begin{theorem}[Canonical equivariant representation]\label{Cer}
There exists a \textit{unique} $($up to isomorphism$)$
N-dimensional irreducible representation $(\Pih,\Hh)$ of $\Ad$ for
which its equivalence class is fixed by $\G$.
\end{theorem}
This means that:
\[
     \Pih  \iso  \Pih^B
\]
for all $B \in \G$.
\\
\\
Since the canonical representation $(\Pih,\Hh)$ is irreducible, by
Schur's lemma we get the canonical projective representation of
$\G$ compatible with $\Pih$:
\begin{corollary}[Canonical projective representation]\label{prq}
There exists a unique projective representation $\rhop: \G \lto
\mathrm{PGL}(\Hh)$ s.t.:
\begin{equation*}
{\rhop (B)}^{-1} \Pih (f) \rhop (B) = \Pih (f\circ B)
\end{equation*}
for all $f \in {\cal A}$ and $B \in \G$.
\end{corollary}
\textbf{Remark.} Corollary \ref{prq} is an improvement to the
known constructions (cf. \cite{HB,Me,KR1}) which has the group $\G
_\Theta := \{\left(
\begin{smallmatrix}
 a & b \\
 c & d
\end{smallmatrix}\right)  | \r ab = cd = 0 \r (2)\}$ as the group of quantizable
elements.
\\
\\
Using a result of Coxeter-Moser \cite{CM} about the structure of
the group $\G$ we get:
\begin{theorem}[Linearization]\label{lin}
The projective representation $\rhop$ can be lifted to a linear
representation in exactly $12$ different ways.
\end{theorem}
\textbf{Remark.} The existence of the linear representation
$\rhoh$ in Theorem \ref{lin} answers Hannay-Berry's question (cf.
\cite{HB, Me}) on the multiplicativity of the map $\rhoh$.
\\
\\
\textbf{Summary.} For $\h \in \Q$ let $(\rhoh,\Pih,\Hh)$ be the
canonical (projective) equivariant quantization of $\T$. We can
endow the space $\Hh$ with a canonical unitary structure s.t.
$\Pih$ is a $*$-representation and $\rhoh$ is unitary. This
``family'' of $*-$representations of $\Ad$ is by definition a Weyl
quantization of the functions on the torus. The above results show
the existence of a canonical projective equivariant quantization
of the torus, and the existence of a linear equivariant
quantization of the torus.
\subsection*{Acknowledgments}
We would like to thank our Ph.D. adviser J. Bernstein for his
interest and guidance in this project. We very much appreciate the
help of P. Kurlberg and Z. Rudnick who discussed with us their
papers and explained their results. This paper was written during
our visit to MSRI, University of California Berkeley and IHES in
the year 2001. We would like to thank these institutions for
excellent working conditions.
\section{Construction}
We consider the algebra ${\cal A} := \F$ of smooth complex valued
function on the torus and the dual lattice $\Lam ^* := \{ \xi \in
V^* | \; \xi(\Lam) \subset \Z \} $. Let $<,>$ be the pairing
between $\W$ and $\W^*$. The map $\xi \mapsto s(\xi)$ where
$s(\xi)(x) := e^{2\pi i <x,\xi>},\r x\in \T$ and $\xi\in \Lam^*$
defines a canonical isomorphism between $\Lam^*$ and the group
$\Td  := \mathrm{Hom} (\T,\C^\ast)$ of characters of $\T$.

\subsection{The quantum tori}
Fix $\h \in \R.$ The Rieffel's quantum torus (cf. \cite{Ri}) is
the non-commutative algebra $\Ad$ defined over $\C$ by generators
$\{s(\xi),\r \xi \in \Lam ^* \}$, and relations:
\begin{equation*}
s(\xi+\eta) = e^{\pi i \h \ome(\xi,\eta)} s(\xi)s(\eta)
\end{equation*}
for all $\xi,\eta \in \Lam^*$.
\\
\\
Note that the lattice $\Lam^*$ serves, using the map $\xi\mapsto
s(\xi)$, as a basis for the algebra $\Ad$. This induces an
identification of vector spaces $\Ad\iso \A$ for every $\h$. We
will use this identification in order to view elements of the
(commutative) space $\A$ as members of the (non-commutative) space
$\Ad$.
\subsection{Weyl quantization}
To get a Weyl quantization of $\cal A$ we use a specific
one-parameter family of representations (see subsection \ref{Ceq}
below) of the quantum tori. This defines an operator $\Pih(\xi)$
for every $\xi \in \Lam^*$. We extend the construction to every
function $f \in {\cal A}$ using the Fourier theory. Suppose:
\begin{equation*}
f = \sum\limits_{\xi \in \Lam^*} a_{_\xi}\cdot\xi
\end{equation*}
is its Fourier expansion. Then we define its {\it Weyl
quantization} by:
\begin{equation*}
\Pih(f) := \sum\limits_{\xi \in \Lam^*} a_{_\xi} \Pih(\xi).
\end{equation*}
The convergence of the last series is due to the rapid decay of
the Fourier coefficients of the function $f$.

\subsection{Projective equivariant  quantization}
The group $\G = \mathrm{SL}_2(\Z)$ acts on $\Lam$ preserving
$\ome$. Hence $\G$ acts on $\Ad$ and the formula of this action is
$s^B(\xi) := s(B\xi)$. Given a representation $(\Pih,\Hh)$ of
$\Ad$ and an element $B \in \G$, define $\Pih^B(s(\xi)) :=
\Pih(s^{B^{-1}}(\xi))$. This formula induces an action of $\G$ on
the set Irr$(\Ad)$ of equivalence classes of irreducible algebraic
representations of $\Ad$.
\begin{lemma}\label{Fdim} All irreducible representations of $\Ad$
are $N$-dimensional.
\end{lemma}
Now, suppose $(\Pih,\Ad,\Hh)$  is an irreducible  representation
for which its equivalence class is fixed by the action of $\G$.
This means that for any $B\in \G$ we have $\Pih \iso \Pih^B$, so
by definition there exists an operator $\rhoh(B) \in \GL(\Hh)$
such that:
\begin{equation*}
{\rhoh(B)}^{-1} \Pih(\xi) \rhoh(B) = \Pih(B\xi)
\end{equation*}
for all $\xi\in \Lam ^*$. This implies the Egorov identity
(\ref{eqp}) for any function . Now, since $(\Pih,\Hh)$ is an
irreducible representation then by Schur's lemma for every $B\in
\G$ the operator $\rhoh(B)$ is uniquely defined up to a scalar.
This implies that $(\rhoh,\Hh)$ is a projective representation of
$\G$.
\subsection{The canonical equivariant quantization}\label{Ceq}
In what follows we consider only the case  $\h \in \Q$. We write
$\h$ in the form $\h = \frac{M}{N}$ with $\mathrm{gcd}(M,N) =1$.
\begin{proposition}\label{uf}
There exists a unique $\Pih \in \Irr$ which is a fixed point for
the action of $\G$.
\end{proposition}
\subsection{Unitary structure}
Note that $\Ad$ becomes a $\r *-$ algebra using the formula
$s(\xi)^* := s(-\xi)$. Let $(\Pih,\Hh)$ be the canonical
representation of $\Ad$.
\begin{remark}
There exists a canonical $($unique up to scalar$)$ unitary
structure on $\Hh$  for which $\Pih$ is a $*-$representation.
\end{remark}
\subsection{Realization}
Choosing a symplectic basis for $\Lam^*$ we get the
identifications $\Lam ^* \iso \Z \oplus \Z$ and $\G =$ SL$_2
(\Z)$. We will consider the realization on the Hilbert space:
$$
\H := \mathrm{L}^2(\Z/N\Z).
$$
\subsubsection{Formula for $\pi$} The representation $\pi$ is given by:
\begin{equation*}
\left[\pi(\m,\n)\f\right](\x) = \alpha(\m,\n) \psi(\n\x)\f(\x+\m),
\end{equation*}

where $\alpha (\m,\n) := (-1)^{M(\m+\n)}e^{\pi i \h \m\n}$ and
$\psi(t)$ denote the additive character $\psi(t) := e^{2 \pi i \h
t}$ on $\Z/N\Z$.
\subsubsection{formula for $\rho$} The projective representation $\rho$
is described by the following formulas:

\begin{eqnarray*}
\left[\rho\left(
\begin{array}{cc}
 1 & 1 \\
 0 & 1
\end{array} \right)\f\right](\x) & = & \mathrm{\texttt{Q}}(\x)\f(\x),
\end{eqnarray*}
where $\texttt{Q}(\x) := (-1)^{\varepsilon\x}e^{\pi i \h \x^2}$,
with $\varepsilon := MN$ (mod 2), and:
\begin{eqnarray*}
\left[\rho\left(
\begin{array}{cc}
 0 & -1 \\
 1 & 0
\end{array} \right)\f\right](\x) & = & \widehat{\f}(\x),
\end{eqnarray*}
where $\widehat{\f}$ denote the Fourier transform:
$$
\widehat{\f}(\x) := \frac{1}{\sqrt{N}} \sum\limits_{\y \in \Z/NZ}
\f(\y) \psi (\y\x).
$$
\section{Proofs}
\subsection{Proof of Lemma \ref{Fdim}} Suppose $(\Pih,\Hh)$ is an irreducible representation of $\Ad$.
\\
\\
\textbf{Step 1.} First we show that $\Hh$ is finite dimensional.
$\Ad$ is a finite module over $\cent (\Ad) = \{s(N\xi),\r \xi \in
\Lam ^* \}$ which is contained in the center of $\Ad$. Because
$\Hh$ has at most countable dimension (as a quotient space of
$\Ad$) and $\C$ is uncountable then by Kaplansky's trick (cf.
\cite{MR}) $\cent (\Ad)$ acts on $\Hh$ by scalars. Hence dim $\Hh
< \infty $.
\\
\\
\textbf{ Step 2.} We show that $\Hh$ is N-dimensional. Choose a
basis $(e_1,e_2)$ of $\Lam ^*$ s.t. $\ome(e_1,e_2)=1$. Suppose
$\lam \neq 0$ is an eigenvalue of $\Pih (e_1)$ and denote by $\H
_\lam$ the corresponding eigenspace. We have the following
commutation relation $\Pih (e_1) \Pih (e_2) = \gamma  \Pih (e_2)
\Pih (e_1)$  where $\gamma := e^{-2 \Pi i \frac{M}{N}}$. Hence
$\Pih (e_2): \H _{\gamma ^j \lam} \lto \H _{\gamma ^{j+1} \lam},$
and because $\mathrm{gcd}(M,N) = 1$ then $\r \H _{\gamma ^i \lam}
\neq \H _{\gamma ^j \lam} $ for $ 0\leq i\neq j \leq N-1$. Now,
let $v \in \H _\lam$ and recall that $\Pih (e_2) ^N = \Pih(Ne_2)$
is a scalar operator. Then the space $\mathrm{span}\{ v,\Pih
(e_2)v,\ldots,\Pih (e_2) ^{N-1} v\}$ is N-dimensional $\Ad
-$invariant subspace hence it equals $\r \Hh$. $\EProof$
\subsection{Proof of Proposition \ref{uf}}
Let us show the existence of a unique fixed point for the action
of $\G$ on Irr$(\Ad)$.
\\
\\
Suppose $(\Pih,\Hh)$  is an irreducible representation of $\Ad$.
By Schur's lemma for every $\xi \in \Lam^*$ the operator
$\Pih(N\xi)$ is a scalar operator, i.e., $\Pih (N\xi) = q
_{_{\Pih}}(\xi) \cdot \mathrm{I}$. We have $\Pih(0) = \mathrm{I}$
and hence $q _{_{\Pih}}(\xi)\neq 0$ for all $\xi \in \Lam^*$. Thus
to any irreducible representation we have attached a scalar
function $q _{_{\Pih}}:\Lam ^* \lto \C^*$. Consider the set $Q
_{_\h}$ of \textit{twisted characters} of $\Lam ^*$:
\begin{equation*}
Q _{_\h} := \{ q:\Lam ^* \lto \C ^* , \r q(\xi+\eta) = (-1)^{M N
w(\xi,\eta)} q(\xi)q(\eta) \}.
\end{equation*}
The group $\G$ acts naturally on this space by
$q^B(\xi):=q(B^{-1}\xi)$. It is easy to see that we have defined a
map $\q:\Irr \lto Q _{_\h}$ given by $\Pih \mapsto q _{_{\Pih}}$
and it is obvious that this map is compatible with the action of
$\G$. We use the space of twisted characters in order to give a
description for the set Irr$(\Ad)$:
\begin{lemma}\label{realizati}
The map $\Pih  \mapsto q _{_{\Pih}}$ is a  $\G$-equivariant
bijection:
\[
\begin{CD}
\q : \mathrm{Irr}(\Ad)    @>  >>  Q _{_\h}.
\end{CD}
\]
\end{lemma}
Now, Proposition \ref{uf} follows from the following claim:
\begin{claim}\label{ufpoint}
There exists a unique $q _{_o} \in Q _{_\h}$ which is a fixed
point for the action of $\G$.
\end{claim}
\textbf{Proof of Lemma \ref{realizati}.} \textbf{Step 1.} The map
$\q$ is surjective. Denote by $\AT:= \mathrm{Hom} (\Lam ^*,\C ^*)
\r $ the group of complex characters of $\Lam ^*$. We define an
action of $\AT$ on $\Irr$ and on $Q _{_\h}$ by $\Pih \mapsto
\chi\Pih$ and $q \mapsto \chi^N q$, where $\chi \in \AT$, $\Pih
\in \Irr$ and $q \in Q _{_\h}$. The map $\q$ is clearly a
$\AT$-equivariant map with respect to these actions. Since $\q$ is
$\AT$-equivariant, it is enough to show that the action of $\AT$
on $Q _{_\h}$ is transitive.
 Suppose $q _{_1} , q _{_2} \in Q _{_\h}$. By definition there exists a character $\chi _{_1} \in \AT$
 for which $\chi _{_1} q _{_1}  = q _{_2}$. Let
$\chi$ be one of the $N$'s roots of  $\chi _{_1}$ then $\chi ^N q
_{_1} = q _{_2}$.
\\
\\
\textbf{Step 2.} The map $\q$ is one to one. Suppose $(\Pih,\Hh)$
is an irreducible representation of $\Ad$. It is easy to deduce
from the proof of Lemma \ref{Fdim} (Step 2) that for $\xi \notin
N\Lam ^*$ we have $\mathrm{tr}(\Pih(\xi)) = 0$. But we know from
character theory that an isomorphism class of a finite dimensional
irreducible representation of an algebra is recovered from its
character. This completes the proof of Lemma \ref{realizati}.
$\EProof$
\\
\\
\textbf{Proof of Claim \ref{ufpoint}.} {\it Uniqueness}. Fix $q
\in Q _{_\h}$. The map $\chi \mapsto \chi q$ give a bijection of
$\AT$ with $Q _{_\h}$. But the trivial character $\textbf{1} \in
\AT$ is the unique fixed point for the action of $\G$ on $\AT$.
\\
{\it Existence.} Choose a basis $(e_1,e_2)$  of $\Lam ^*$ s.t.
$\ome(e_1,e_2) = 1$. This allows to identify $\Lam ^*$ with $\Z
\oplus \Z$. It is easy to see that the function:
\begin{equation*}
q _{_o}(m,n) =(-1)^{MN(mn+m+n)}
\end{equation*}
is a twisted character which is fixed by $\G$. This completes the
proof of Claim \ref{ufpoint} and of Proposition \ref{uf}.
$\EProof$
\\
\\
\subsection{Proof of Theorem \ref{lin}} The theorem follows from
the following proposition:
\begin{proposition}
Fix a projective representation $\rhop : \G \lto
\mathrm{GL}(\Hh)$. Then it can be lifted to a linear
representation in exactly $12$ ways.
\end{proposition}
\BProof Existence. We want to find constants $c(B)$ for every $B
\in \G$ s.t. $\rhoh := c(\cdot) \rhop$ is a linear representation
of $\G$. This is possible to carry out due to the following fact:
\begin{lemma}[\cite{CM}]\label{present}
The group $\G$ is isomorphic to the group generated by three
letters $S,\; B$ and $Z$ subjected to the relations: $Z^2 = 1$ and
$\r S^2 = B^3 = Z$.
\end{lemma}
Lemma \ref{present} $\Rightarrow$ Existence. We need to find
constants $\r c_{_Z}, c_{_B}, c_{_S} \r $ so that the operators
$\rhoh (Z) := c_{_Z} \rhop (Z),\r \rhoh (B) := c_{_B} \rhop (B),\r
\rhoh (S) := c_{_S} \rhop (S)$ will satisfy the identities:
\begin{equation*}
\rhoh (Z) ^2 = I, \r \rhoh (B) ^3 = \rhoh (Z),\r \rhoh (S) ^2  =
\rhoh (Z).
\end{equation*}
This can be done by taking appropriate scalars.
\\
\\
Now, fix one lifting $\rho_{_0}$. Then for the collection of
operators $\rhoh(B)$ which lifts $\rhop$ define a function
$\chi(B)$ by $\rhoh(B) = \chi(B) \rho_{_0}(B)$. It is obvious that
$\rhoh$ is a representation if and only if $\chi$ is a character.
Thus liftings corresponds to characters. By Lemma \ref{present}
the group of characters $\G^\vee := \mathrm{Hom}(\G,\C^*)$ is
isomorphic to $\Z/12\Z$.$\EProof$.


\begin{thebibliography}{99}
{\small {
\bibitem[HB]{HB} Hannay J.H. and Berry M.V., Quantization of linear maps on the
torus - Fresnel diffraction by a periodic grating. {\it Physica
D1} (1980), 267-291.
%
\bibitem[CM]{CM} Coxeter H. S. M. and Moser, W. O. J.,
Generators and relations for discrete groups. {\it Ergebnisse der
Mathematik und ihrer Grenzgebiete, 14. Springer-Verlag, Berlin-New
York} (1980).
%
\bibitem[D]{D} Degli Esposti M., Quantization of the orientation preserving
automorphisms of the torus. {\it Ann. Inst. Henri Poincare 58}
(1993), 323-341.
%
\bibitem[D4]{D4} Deligne P., Note on quantization. {\it Quantum fields
   and strings: a course for mathematicians, Vol. 1, 2 (Princeton, NJ,
   1996/1997),  Amer. Math. Soc. Providence, RI}  (1999)  367--375.
%
\bibitem[KR1]{KR1} Kurlberg P. and Rudnick Z., Hecke theory and
equidistribution for the quantization of linear maps of the torus.
{\it Duke Math. Jour. 103} (2000), 47-78. Series, 33. Princeton
University Press, Princeton, N.J.} (1980).
%
\bibitem[Me]{Me} F. Mezzadri, On the multiplicativity of quantum cat maps. {\it Nonlinearity 15} (2002), 905-922.
%
\bibitem[MR]{MR} McConnell J.C. and Robson J.C., Non-commutative Noetherian Rings. {\it Graduate studies in
mathematics, 30. Amer. Math. Soc. Providence, RI} (2001), 343-344.
%
\bibitem[Ri]{Ri} Rieffel M.A., Non-commutative tori---a case study of
non-commutative differentiable manifolds, {\it Contemporary Math.
105} (1990), 191-211. }
\end{thebibliography}
\end{document}